\documentclass[12pt]{article}
\usepackage{graphicx}
\usepackage{subfig}

\newcommand\pubnumber{}
\newcommand\pubdate{\today}

\def\cpt{$^a$Centre de Physique Th\'eorique\support. CNRS Luminy, Case 907,
  F-13288 Marseille Cedex 9, France.}

\def\wup{$^b$Bergische Universit\"at Wuppertal, Gaussstr. 20, D-42119
  Wuppertal, Germany.} 

\def\jul{$^c$J\"ulich Supercomputing Center, Forschungszentrum
  J\"ulich, D-52425 J\"ulich, Germany.}  

\def\bud{$^d$Institute for Theoretical Physics, E\"otv\"os University,
  H-1117 Budapest, Hungary.}

\def\support{\footnote{CPT is research unit
UMR 6207 of the CNRS and of the universities Aix-Marseille I, Aix-Marseille II
and Sud Toulon-Var, and is affiliated with the FRUMAM.}}

\def\Title#1{\begin{center} {\Large #1 } \end{center}}
\def\Author#1{\begin{center}{ \sc #1} \end{center}}
\def\Address#1{\begin{center}{ \it #1} \end{center}}

\newcommand\pubblock{\rightline{\begin{tabular}{l} \pubnumber\\
         \pubdate  \end{tabular}}}
\newenvironment{Abstract}{\begin{quotation}  }{\end{quotation}}
\newenvironment{Presented}{\begin{quotation} \begin{center} 
             PRESENTED AT\end{center}\bigskip 
      \begin{center}\begin{large}}{\end{large}\end{center} \end{quotation}}

\def\beq{\begin{equation}}
\def\eeq#1{\label{#1}\end{equation}}
\def\eeqn{\end{equation}}

\def\beqa{\begin{eqnarray}}
\def\eeqa#1{\label{#1}\end{eqnarray}}
\def\eeqan{\end{eqnarray}}

\let\bar=\overbar

\def\Dslash{\not{\hbox{\kern-4pt $D$}}}
\def\dslash{\not{\hbox{\kern-2pt $\del$}}}

\def\msb{{\bar{\ssstyle M \kern -1pt S}}}

\def\lsim{\raise0.3ex\hbox{$<$\kern-0.75em\raise-1.1ex\hbox{$\sim$}}}
\def\gsim{\raise0.3ex\hbox{$>$\kern-0.75em\raise-1.1ex\hbox{$\sim$}}}

\begin{document}
\begin{titlepage}
\pubblock

\vfill
\Title{$F_K/F_\pi$ from the Budapest-Marseille-Wuppertal Collaboration}
\vfill
\Author{A. Ramos$^a$,  S.\,D\"urr$^{b,c}$, Z.\,Fodor$^{b,c,d}$, 
  C.\,Hoelbling$^b$, S.\,D.\,Katz$^{b,d}$, S.\,Krieg$^b $,
  T.\,Kurth$^b$, L.\,Lellouch$^a$, T.\,Lippert$^{b,c}$,
  K.\,K.\,Szab\'o$^b$
  \\[4mm]
  [Budapest-Marseille-Wuppertal Collaboration]
}
\Address{\cpt}
\Address{\wup}
\Address{\jul}
\Address{\bud}
\vfill
\begin{Abstract}
Based on a series of lattice calculations we determine the ratio 
$F_K/F_\pi$ in QCD. With experimental data from kaon decay and nuclear
double beta decay, we obtain a precise determination of
$|V_{us}|$. Our simulation includes $2+1$ flavours of sea quarks, with
three lattice spacings, large volumes and a simulated pion mass
reaching down to about 190\,MeV for a full control over the systematic 
uncertainties. 
\end{Abstract}
\vfill
\begin{Presented}
Proceedings of CKM2010, the 6th International Workshop on the CKM
Unitarity Triangle, University of Warwick, UK, 6-10 September 2010. 
\end{Presented}
\vfill
\end{titlepage}
\def\thefootnote{\fnsymbol{footnote}}
\setcounter{footnote}{0}

\section{Introduction}

The measurement of CKM matrix elements involves examining weak decays of
hadrons. The theoretical description of these decays involves
computing strongly interacting effects of quarks and gluons inside the
hadrons. These effects should be determined in a
fully non-perturbative way. Any weak decay can be written as the
product of three factors 
\begin{equation}
  \textrm{Weak decay} = \{ \textrm{Kinematic}\}\times 
  \{\textrm{CKM element}\}\times 
  \{\textrm{Non-perturbative QCD}\}.
\end{equation}

Computing the non-perturbative QCD effects is a difficult task. A first
principle analytical approach does not yet exists. We have to live
either with numerical simulations or with models of QCD. 

Here we will use lattice QCD as a numerical tool to compute non
perturbative QCD meson decay contants.

\subsection{Light meson decay constants and $|V_{us}|$}

We will follow a proposal by Marciano~\cite{Marciano:2004uf} of
examining the ratio of decay widths
\begin{equation}
  {\Gamma(K\to \mu\bar\nu_\mu)\over\Gamma(\pi\to \mu\bar\nu_\mu)}
=
{|V_{us}|^2\over|V_{ud}|^2}
\frac{F_K^2}{F_\pi^2}
{M_K(1-m_\mu^2/M_K^2)^2\over M_\pi(1-m_\mu^2/M_\pi^2)^2}
\left\{
1+{\alpha\over\pi}(C_K-C_\pi)
\right\}
\label{marciano}
\end{equation}
where this ratio, even including
electromagnetic corrections, are experimentally know with 
an $0.4\%$ precision. Masses are known at least with a relative
precision of $3\!\times\!10^{-5}$. This implies that an accurate
determination of the ratio $F_K/F_\pi$ can be used to determine
$|V_{us}|/|V_{ud}|$. The most recent update of the Flavianet 
kaon working group is~\cite{Antonelli:2008aa}:
\begin{equation}
  {|V_{us}|\over |V_{ud}|}\,{F_K\over F_\pi}=0.27599(59)\;.
\end{equation}
This may be used with $|V_{ud}|=0.97425(22)$~\cite{Hardy:2008gy} from
super-allowed nuclear $\beta$ decays and the lattice determination of
$F_K/F_\pi$ to obtain $|V_{us}|$.

\section{Simulation details}

\begin{figure}[t,b]
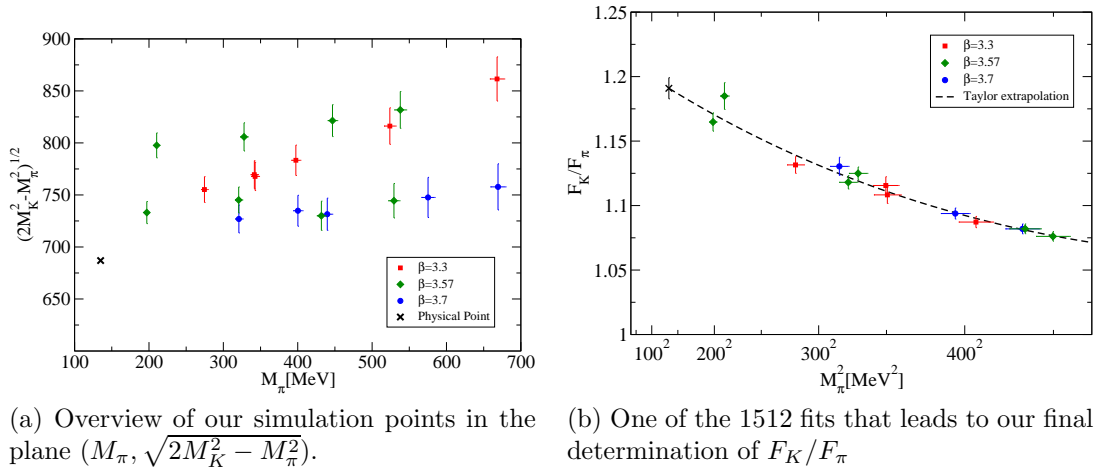

  \centering
  \subfloat[Overview of our simulation points in the plane $(M_\pi,
    \sqrt{2M_K^2-M_\pi^2})$.]{\label{fig:landscape}\includegraphics[width=7.0cm]{fig/landscape}}\quad 
  \subfloat[One of the 1512 fits that leads to our final determination
  of
  $F_K/F_\pi$]{\label{fig:fkfpi}\includegraphics[width=7.0cm]{fig/fkfpi}}\\
  \caption{Overview of simulation points and a representative fit.}
\end{figure}

We simulate 2 degenerate light clover fermions representing the $u$
and $d$ quarks in the strong isospin limit plus a third heavier clover
fermion 
that represents the strange quark. Whereas the value of the strange
quark mass is held fixed close to its physical value light quark
masses are 
varied through a significant range so that our $\pi$ mesons have
masses in the range $190\textrm{ MeV}\lsim M_\pi
\lsim 460\textrm{ MeV}$. This allows a controlled extrapolation to the
physical point $M_\pi \approx 135\textrm{ MeV}$ (see
Figure~\ref{fig:landscape}). 

We simulate at three different lattice spacings
$a\!\simeq\!0.124\textrm{ fm}$ $(\beta=3.3)$,
$a\!\simeq\!0.083\textrm{ fm}$ 
  $(\beta=3.57)$, and $a\!\simeq\!0.065\textrm{ fm}$ $(\beta=3.7)$ for
  a controlled extrapolation to the continuum limit. 

All our simulations are done in large volumes $M_\pi L \gsim 4$, so
that finite volume corrections are below $1\%$. 

More details about both our fermion and gauge actions can be found
in~\cite{Durr:2008rw}. 

\section{Systematic uncertainties.}

To obtain a QCD results from lattice data one has to perform several
extrapolations: one has to extrapolate data to the physical
value of the quark masses (chiral extrapolation), one has to
to the infinite volume limit, and to the
continuum limit. One has also to deal with the contribution from exited
states and the possible uncertainty associated with determining the
QCD coupling (i.e. setting the scale). All these considerations should
be taken into account and each of them should be reflected in the
final uncertainty of our result.  

A detailed description of the method sketched here can be found in the
original work~\cite{Ramos:fkfpip}. 

\subsection{Scale setting issues}

$2+1$ flavour QCD has three independent parameters: the light quark
masses, the strange quark mass, and the gauge coupling. One needs
three experimental inputs to fix these parameters in our simulations. 

To fix quark masses one uses the experimental values of the masses of
the $\pi$ and K mesons, $M_\pi$ and $M_K$.

To set the lattice spacing, one uses a physical state whose mass
does not strongly depend on quark masses. We used both the $\Omega$
and $\Xi$ states, the differences between these two choices measures
the error associated with our choice of physical input.

All physical inputs are corrected to take into account
that we simulate in the isospin symmetric limit without electromagnetic 
interactions. 

\subsection{Chiral extrapolation}

There exist basically two approaches that can guide us in the
extrapolation to the physical point. We can use an effective field
theory approach ($\chi$PT), either in the two or three flavour
setups. 

Another possibility comes from the observation that physical
quantities like decay constants or masses have an analytical expansion
in the quark masses as long as this expansion performed around a
regular point far from the singular point $m_q = 0$. If our lattice
data is close enough to the physical point these analytical expansions
should be a fair approximation to the behaviour of the physical
quantity up to the physical point.

We will fit our data to a total of 7 different
expressions that correspond to one of the previously mentioned
approaches. 

For the case of three flavour $\chi$PT our data is well fitted using
the NLO relation for the ratio $F_K/F_\pi$~\cite{Gasser:1984ux}. In
fact up to higher order 
terms, the NLO relation can be written in three different ways. The
difference between these three expressions measures the contribution
of higher order contributions.

The ratio $F_K/F_\pi$ can be obtained from the pion mass dependence of
$F_K$ in the heavy kaon two flavour $\chi$PT
formalism and the pion mass dependence of $F_\pi$
in $SU(2)$ $\chi$PT~\cite{Allton:2008pn}. Our data is well fitted by
the NLO relation, and up to higher order terms, we have two different
ways of expanding the ratio of decay constants. The difference between
these two expressions are higher order contributions. 

Finally the ratio of decay constants have a perfectly regular
expansion, as soon as this expansion is done far from the singular
point $m_q = 0$. In fact our lattice data is well fitted by a simple
polynomial expansion around a regular point that lies midway between
the physical point and our heaviest pion ensemble. One can also fit
the data to a rational approximation, that can be seen as a Pad\'e
like ansatz. This gives a total of 2
expressions based on an analytical expansion around a regular point
that, again, differ in higher order terms.

It is important to remark that the total of 7 formulae differ both in
the approach (three, two flavour $\chi$PT or analytical expansions),
and in the higher order contributions within a framework (NNLO
contributions in $\chi$PT or higher powers in the analytical
expansions.)

\subsection{Continuum limit}

The ratio $F_K/F_\pi$ is a flavour breaking term. This means that no
matter the value of the lattice spacing we will always have
$F_K/F_\pi=1$ when $m_{ud} = m_s$. This suggests that cutoff effects
should be $\propto (M_\pi^2-M_K^2)$. 

Our action is formally only improved up to order $\mathcal O(\alpha_s
a)$, but we have numerical evidence~\cite{Durr:2008rw} that cutoff
effects scale as $\mathcal O(a^2)$. 

In particular for the present case, cutoff effects are small. This can
be understood by 
looking at $SU(3)$ $\chi$PT, where the leading correction cancels for the
ratio of decay constants. For our particular dataset they are
statistically consistent with zero (i.e. the statistical uncertainty
of our data are bigger than lattice spacing corrections.)

Taking these considerations into account we choose to parametrise
cutoff effects in our data by adding or not a term
\begin{equation}
\left.\frac{F_K}{F_\pi}\right\vert_\textrm{c.o.} =c\times\left\{
\begin{array}{l}
a(M_K^2-M_\pi^2)/\mu_\textrm{QCD}\\
\quad\textrm{or}\\
a^2(M_K^2-M_\pi^2)\\
\end{array}
\right.
\label{eq:su3co}
\end{equation}
to any of the previous 7 formulae that gives the dependence of
$F_K/F_\pi$ with meson masses.

\subsection{Finite volume corrections}

For large enough lattices, finite volumes effects scale as $\propto
e^{-M_\pi L}$. All our simulation points respect the bound $M_\pi
\gsim 4$, making finite volume corrections small. In fact our data is
well fitted without any kind of finite volume correction, indicating
that the size of the correction is smaller than the
statistical accuracy of our data. 

Nevertheless, corrections due to finite volume are known in
$\chi$PT. Meson masses and decay constants receive corrections that
have been computed. We decided to correct the decay constants with a 2
loop expression~\cite{Colangelo:2005gd}. To estimate the uncertainty
of this correction, we 
use the 1 loop expression together with an upper bound on the
correction (coming from correcting only $F_\pi$). 

\subsection{Excited state contributions}

Masses and decay constants are obtained from the large time
behaviour of correlators, i.e.
\begin{equation}
  C(t)\equiv\left(\frac{a}{L}\right)^3\sum_{\vec{x}}\langle[\bar
  d\gamma_5 u](x) 
  [\bar u\gamma_5 d](0)\rangle\stackrel{0\ll t\ll T}{\longrightarrow}
  \frac{\langle 0|\bar d\gamma_5 u|\pi^+(\vec{0})\rangle
    \langle \pi^+(\vec{0})|\bar u\gamma_5 d|0\rangle}{2M_\pi}\,e^{-{
      M_\pi} t}.
\end{equation}

We choose for each value of the coupling $\beta$ the start of the
fitting interval $t_{\textrm{min}}$
so that it is strongly dominated by the ground state. There are
several reasonable possibilities for this, and we choose to use 
18 possibilities that will receive different tiny contributions
from excited states. These different fitting intervals will be used to
estimate the uncertainties due to excited state.

\section{Results}

The procedure described above lads to a total of $2\times 18\times
7\times 2\times 3 = 1512$ different ways of obtaining the ratio
$F_K/F_\pi$ in the continuum, at the physical mass point and in the
infinite volume limit (see Figure~\ref{fig:fkfpi}.) 

Since not all the previous procedures describes our data equally well
(although the $\chi^2/\textrm{dof}$ for all the fits is close to 1),
we weight each of the values with the quality of the fit to produce a
distribution of values.

The median, typical result of our analysis,
is taken as our final result. The 16-th/84-th percentiles, that
measures the ``spread'' of values, give our final systematic
uncertainty. To determine the statistical accuracy of our results, we
resample our configurations 2000 times, and the standard deviation of
the median over these 2000 samples is taken as the statistical
uncertainty.  

Our final result is
\begin{equation}
  {F_K\over F_\pi}\bigg|_\textrm{phys}=1.192(7)_\textrm{stat}(6)_\textrm{syst}
\end{equation}
Using the value $|V_{ud}| = 0.97425(22)$\cite{Hardy:2008gy} we obtain 
\begin{equation}
  |V_{us}| = 0.2256(18). 
\end{equation}
Using also $|V_{ub}| = (3.93\pm 0.36)\times
10^{-3}$ the first row CKM unitarity relation is
\begin{equation}
  |V_{ud}|^2+|V_{us}|^2+|V_{ub}|^2 = 1.0001(9)
\end{equation}
with no signal for physics beyond the standard model. For details
on the calculations see~\cite{Ramos:fkfpip}.

\bibliography{/home/alberto/latex/math,/home/alberto/latex/campos,/home/alberto/latex/fisica,/home/alberto/latex/computing}

\begin{thebibliography}{1}

\bibitem{Marciano:2004uf}
William~J. Marciano.
%\newblock {Precise determination of |V(us)| from lattice calculations of
%  pseudoscalar decay constants}.
\newblock {\em Phys. Rev. Lett.}, 93:231803, 2004.

\bibitem{Antonelli:2008aa}
M.~Antonelli et~al.
%\newblock ({FLAVIAnet} working group on kaon decays).
\newblock {\em Nucl. Phys. B Proc. Suppl.}, 81:181--183, 2008.

\bibitem{Hardy:2008gy}
J.~C. Hardy and I.~S. Towner.
%\newblock {Superallowed 0+ to 0+ nuclear beta decays: A new survey with
%  precision tests of the conserved vector current hypothesis and the standard
%  model}.
\newblock {\em Phys. Rev.}, C79:055502, 2009.

\bibitem{Durr:2008rw}
S.~Durr et~al.
%\newblock {Scaling study of dynamical smeared-link clover fermions}.
\newblock {\em Phys. Rev.}, D79:014501, 2009.

\bibitem{Ramos:fkfpip}
S.~Durr et~al.
%\newblock {The ratio FK/Fpi in QCD}.
\newblock {\em Phys. Rev.}, D81:054507, 2010.

\bibitem{Gasser:1984ux}
J.~Gasser and H.~Leutwyler.
%\newblock {Low-Energy Expansion of Meson Form-Factors}.
\newblock {\em Nucl. Phys.}, B250:517--538, 1985.

\bibitem{Allton:2008pn}
C.~Allton et~al.
%\newblock {Physical Results from 2+1 Flavor Domain Wall QCD and SU(2) Chiral
%  Perturbation Theory}.
\newblock {\em Phys. Rev.}, D78:114509, 2008.

\bibitem{Colangelo:2005gd}
Gilberto Colangelo, Stephan Durr, and Christoph Haefeli.
%\newblock {Finite volume effects for meson masses and decay constants}.
\newblock {\em Nucl.Phys.}, B721:136--174, 2005.

\end{thebibliography}

\end{document}